\documentclass[letterpaper, 10 pt, conference]{ieeeconf}
\IEEEoverridecommandlockouts     
\usepackage{graphics} 
\usepackage{svg}
\usepackage{amsmath} 
\usepackage{amssymb}
\usepackage{float}
\usepackage{graphicx}
\usepackage[normalem]{ulem}
\usepackage{algorithm}
\usepackage{cite}
\usepackage{algpseudocode}

\usepackage{soul}
\usepackage{comment}
\usepackage{tikz}
\usepackage{pgfplots}
\usepackage{pgfplotstable}
\usepackage{graphviz}
\usetikzlibrary{math}
\usetikzlibrary{shapes,arrows}
\usetikzlibrary{arrows.meta}
\tikzset{>={Latex[width=2.5mm,length=2.5mm]}}
\usetikzlibrary{positioning}
\tikzstyle{block}=[draw opacity=0.7,line width=1.4cm]

\newcommand{\real}{{\mathbb{R}}}

 \newcommand{\boxend}{\hfill \ensuremath{\Box}}
\newtheorem{thm}{Theorem}[section]

\newtheorem{lem}{Lemma}[section]

\newtheorem{prob}{Problem}

\allowdisplaybreaks
\title{\LARGE \bf
Optimality Gap of Decentralized Submodular Maximization under Probabilistic Communication
}
\author{Joan Vendrell and Solmaz Kia, \emph{Senior Member, IEEE} 
\thanks{This work is supported by UCI-LANL fellowship. Authors are with the Mechanical and Aerospace Engineering Department of University of California Irvie, Irvine, CA, USA
        {\tt\small jvendrel,solmaz@uci.edu}
        }
}
\allowdisplaybreaks
\newcommand{\longthmtitle}[1]{\mbox{}\textit{{(#1):}}}
\newcolumntype{d}[1]{>{\centering\arraybackslash}m{#1\linewidth}}
\pdfoutput=1
\pgfplotsset{compat=1.18}
\begin{document}
\maketitle
\begin{abstract}
This paper considers the problem of decentralized submodular maximization subject to partition matroid constraint using a sequential greedy algorithm with probabilistic inter-agent message-passing. We propose a communication-aware framework where the probability of successful communication between connected devices is considered. Our analysis introduces the notion of the probabilistic optimality gap, highlighting its potential influence on determining the message-passing sequence based on the agent's broadcast reliability and strategic decisions regarding agents that can broadcast their messages multiple times in a resource-limited environment. This work not only contributes theoretical insights but also has practical implications for designing and analyzing decentralized systems in uncertain communication environments. A numerical example demonstrates the impact of our results.
\end{abstract}


\section{Introduction}
\label{sec::intro}
\begin{figure}[t]
    \centering
    \includegraphics[width=0.4\textwidth]{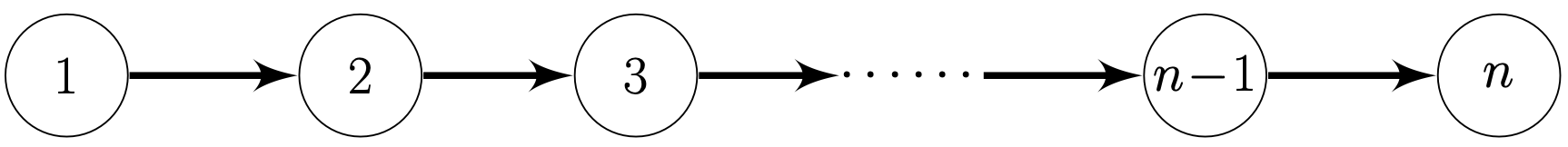}
    \caption{{\small A directed chain graph where an arrow from one node to
another signifies an edge, indicating directional information
flow from the tail node (in-neighbor) to the head node (out-neighbor).}}
\label{fig:chain_graph}
\end{figure}

This paper considers the problem of decentralized submodular maximization subject to a partition matroid when agents implement a \emph{sequential greedy algorithm} with probabilistic inter-agent communication. We study a network of $n$ agents, denoted as $\mathcal{A}=\{1, \ldots, n\}$, each equipped with communication and computation capabilities and interconnected via a directed chain graph (see Fig.~\ref{fig:chain_graph}). The goal for each agent $i \in \mathcal{A}$ is to select up to $\kappa_i \in \mathbb{Z}_{>0}$ strategies from its local distinct discrete strategy set $\mathcal{P}_i$ in way that a normal monotone increasing and submodular utility function \(f:2^\mathcal{P}\to\mathbb{R}_{\geq0}\),  \(\mathcal{P} \!=\! \bigcup_{i \in \mathcal{A}} \mathcal{P}_i\), evaluated at collective strategy set of the agents is maximized. The optimization problem is expressed as

\begin{subequations}\label{eq::mainProblem}
\begin{align}
    &\underset{\mathcal{S}\in \mathcal{I}}{\textup{max}}\,f(\mathcal{S}) \\
    &\mathcal{I} = \big\{ \mathcal{S} \subset \mathcal{P}\,\big|\,\, |\mathcal{S} \cap \mathcal{P}_i|\leq \kappa_i,~ \forall i\in\mathcal{A}\big\},\label{eq::mainProblem_b}
\end{align}
\end{subequations}
\noindent where $\mathcal{I}$ is a \emph{partition matroid}, restricting the number of strategies each agent $i \in \mathcal{A}$ can choose to $\kappa_i$. The agents access the utility function via a black box that returns \(f(\mathcal{S})\) for any set \(\mathcal{S} \subset \mathcal{P}\) (value oracle model).

\setlength{\textfloatsep}{3pt}
\begin{algorithm}[t]
{\footnotesize
\caption{Sequential Greedy Algorithm}\label{alg:sequential}
\begin{algorithmic}
\Require Local sets $\mathcal{P}_1,\cdots,\mathcal{P}_n$ and utility function $f$
\Ensure $\bar{\mathcal{S}}\subset \mathcal{P}$  satisfying $|\bar{\mathcal{S}}\cap \mathcal{P}_i|\leq \kappa_i$, $i \in \mathcal{A}$.
\State $\bar{\mathcal{S}}_{SG} \gets \emptyset$
\State $\bar{\mathcal{S}}_{i} \gets \emptyset$ for $i \in \mathcal{A}$
\For{$i \in \mathcal{A}$}
  \For{$j \in \{1, \dots, \kappa_i\}$}
    \State $s^\star \gets \text{argmax}_{s \in \mathcal{P}_i \backslash \bar{\mathcal{S}}_i}{ \Delta(s|(\bar{\mathcal{S}}_{SG} \cup \bar{\mathcal{S}}_i))}$
    \State $\bar{\mathcal{S}}_i \gets \bar{\mathcal{S}}_{i} \cup \{s^\star\}$
  \EndFor
  \State $\bar{\mathcal{S}}_{SG} \gets \bar{\mathcal{S}}_{SG} \cup \bar{\mathcal{S}}_{i}$
\EndFor
\State $\bar{\mathcal{S}} \gets \bar{\mathcal{S}}_{SG}$
\end{algorithmic}
}
\end{algorithm}

Submodular function maximization problems such as~\eqref{eq::mainProblem} are often NP-hard~\cite{GLN-LAW-MLF:78}. However, submodularity, a property of set functions with deep theoretical implications, enables the establishment of constant factor approximate (suboptimal) solutions for submodular maximization problems. A fundamental result by Nemhauser et al.~\cite{GLN-LAW-MLF:78} establishes that the simple \emph{sequential greedy algorithm} shown in Algorithm~\ref{alg:sequential} is guaranteed to provide a constant $1/2$-approximation factor solution for the submodular maximization problem~\eqref{eq::mainProblem}.

Submodular maximization problems subject to matroid constraints appear in many data-centric problems, such as agglomerative clustering, exemplar-based clustering~\cite{PH-AS-SB-KM:21}, categorical feature compression~\cite{AR-HE-LC-MB-TF-VM:19}, and data subset selection~\cite{KW-RI-JB:15}. To alleviate the burden on central sequential greedy solvers handling massive data, various parallel computing methods such as MapReduce-based solvers or federated learning frameworks~\cite{rafiey2024decomposable} have been proposed.

In the domain of cyber-physical systems, submodular maximization problems of the form~\eqref{eq::mainProblem} appear in various optimal resource allocation and scheduling tasks, such as optimal sensor placement~\cite{NM-RH:18}, energy storage placement~\cite{JQ-IY-RR:19,MB-SP-ED-AV:20}, measurement scheduling~\cite{STJ-SLS:15}, voltage control in smart grids~\cite{ZL-AC-PL-LB-DK-RP:16}, and persistent monitoring via mobile robots~\cite{NR-SSK:21}. These applications often require decentralized solutions, where agents achieve a global optimal solution through local computations and inter-agent interactions/communication. 

Distributed solutions based on continuous relaxations~\cite{AR-AA-BS-JGP-HH:19,NR-SSK:23} are proposed in the literature to solve problem~\eqref{eq::mainProblem} over graph integration topologies, however, they often come with significant communication and computation costs. The sequential decision-making structure of Algorithm~\ref{alg:sequential} lends itself naturally to decentralized implementation over a directed chain graph using sequential message-passing, as described in Algorithm~\ref{alg:sequential_decentral}~\cite{NR-SSK:21}. Starting with the first agent in the chain, each agent, after receiving $\bar{\mathcal{S}}_{SG}$—the compound choices of the preceding agents—from its in-neighbor, runs a local sequential greedy algorithm over its local dataset to make its choices considering the choices already made. After determining its local selections and incorporating them into $\bar{\mathcal{S}}_{SG}$, the agent forwards this updated set to its out-neighbor. This method of sequential message-passing can also be facilitated through cloud access.

\begin{algorithm}[t]
{\footnotesize
\caption{Decentralized Sequential Greedy Algorithm}\label{alg:sequential_decentral}
\begin{algorithmic}
\Require Local sets $\mathcal{P}_1,\cdots,\mathcal{P}_n$ and utility function $f$
\Ensure $\bar{\mathcal{S}}\subset \mathcal{P}$ satisfying $|\bar{\mathcal{S}}\cap \mathcal{P}_i|\leq \kappa_i$, $i\in\mathcal{A}$. 
\State Every agent $i \in \mathcal{A}$ initializes a local copy of $\bar{\mathcal{S}}_{SG} \gets \emptyset$
\State Every agent $i \in \mathcal{A}$ initializes $\bar{\mathcal{S}}_{i} \gets \emptyset$ 
\For{$i \in \mathcal{A}$}
\If{$i \neq 1$}
\State $\bar{\mathcal{S}}_{i} \gets \bar{\mathcal{S}}_{i-1}$
\EndIf
  \For{$j \in \{1,\cdots,\kappa_i\}$}
    \State $s^\star \gets \text{argmax}_{s \in \mathcal{P}_i \backslash \bar{\mathcal{S}}_i}{ \Delta(s | \bar{\mathcal{S}}_{SG} \cup \bar{\mathcal{S}}_i)}$
    \State $\bar{\mathcal{S}}_i \gets \bar{\mathcal{S}}_{i} \cup \{s^\star\}$
  \EndFor
  \State Send $\bar{\mathcal{S}}_{i}$ to agent $i+1$
\EndFor
\State $\bar{\mathcal{S}} \gets \cup_{i \in \mathcal{A}}\bar{\mathcal{S}}_{i} $
\end{algorithmic}}
\end{algorithm}

In practice, message delivery reliability between an in-neighbor and its out-neighbor is not always assured. The effect of delivery failures on a decentralized sequential greedy algorithm has been explored in~\cite{Gharesifard1}; refer to Section~\ref{sec:approach} for more information. This research examines how a specific message passing sequence impacts the optimality gap from a deterministic point of view. However, practical message delivery success is subject to unpredictable factors such as communication channel strength, agent reliability, and network congestion. The impact of probabilistic communication on convergence in various optimization problems is well documented in the literature, such as in~\cite{JP-SW-MJ-KSC:22,rostami2023federated}. These studies offer theoretical insights and have practical significance for the design and analysis of decentralized systems in uncertain communication settings. Considering these uncertainties, this paper addresses the pivotal research question of how probabilistic message-passing affects the guaranteed optimality gap of a decentralized sequential greedy algorithm. We frame this challenge as the following problem statement.

\begin{prob}[Probabilistic Sequential Message-Passing]\label{prob:optimality_gap_prob} Within the framework of the decentralized sequential greedy Algorithm~\ref{alg:sequential_decentral}, assume the probability of a message $\bar{\mathcal{S}}_i$ being successfully transmitted from agent $i$ to its out-neighbor $i+1$ is $p_i$. Determine the probabilistic optimality gap $\alpha_p$ in 
    \begin{equation}\label{eq:expected_optim_gap}
    \mathbb{E}[f(\bar{\mathcal{S}})]\geq \alpha_p \,f(\mathcal{S}^\star),
    \end{equation}
    where $\mathcal{S}^\star$ is the maximizer of optimization problem~\eqref{eq::mainProblem}.\boxend \end{prob}
\medskip

The main objective of this paper is to investigate and address Problem~\ref{prob:optimality_gap_prob}, focusing on characterizing the probabilistic optimality gap, $\alpha_p$. Understanding $\alpha_p$ is crucial for decentralized sequential greedy algorithms operating under probabilistic message-passing conditions. This work not only contributes theoretical insights but also has practical implications for designing and analyzing decentralized systems in uncertain communication environments.

\noindent\emph{Notation and definitions}: For a discrete ground set $\mathcal{P}$, $2^{\mathcal{P}}$ is its power set, the set that contains all the subsets of $\mathcal{P}$. A set function $f: 2^{\mathcal{P}} \rightarrow \real_{ \geq 0}$ is \textit{submodular} if and only if for any $\mathcal{P}_2 \subseteq \mathcal{P}_1 \subseteq {\mathcal{P}}$ and for all $p \in \mathcal{P} \setminus \mathcal{P}_1$ we have that
\begin{equation}
 \label{eqn:diminishing_returns}
 f(\mathcal{P}_2 \cup \{ p \}) - f(\mathcal{P}_2) \geq f(\mathcal{P}_1 \cup \{ p \}) - f(\mathcal{P}_1)
\end{equation}
Function $f$ is normal if $f(\emptyset)=0$ and  is monotone increasing if  for any $\mathcal{P}_1\subset\mathcal{P}_2\subset\mathcal{P}$ we have $ f(\mathcal{P}_1)\geq f(\mathcal{P}_2)$. For any $p\in\mathcal{P} $ and any $\mathcal{P}\subset\mathcal{P}$, $\Delta (p|\mathcal{P})=f(\mathcal{P} \cup \{ p\}) - f(\mathcal{P})$ is the marginal gain of adding $p$ to the set $\mathcal{P}$.

\section{Sequential greedy under probabilistic inter-agent message-passing}
\label{sec:approach}
\begin{figure}[t]
    \centering
    \begin{tabular}{p{0.2\textwidth} @{\hspace{15pt}} p{0.2\textwidth}}
        \includegraphics[width=0.187\textwidth]{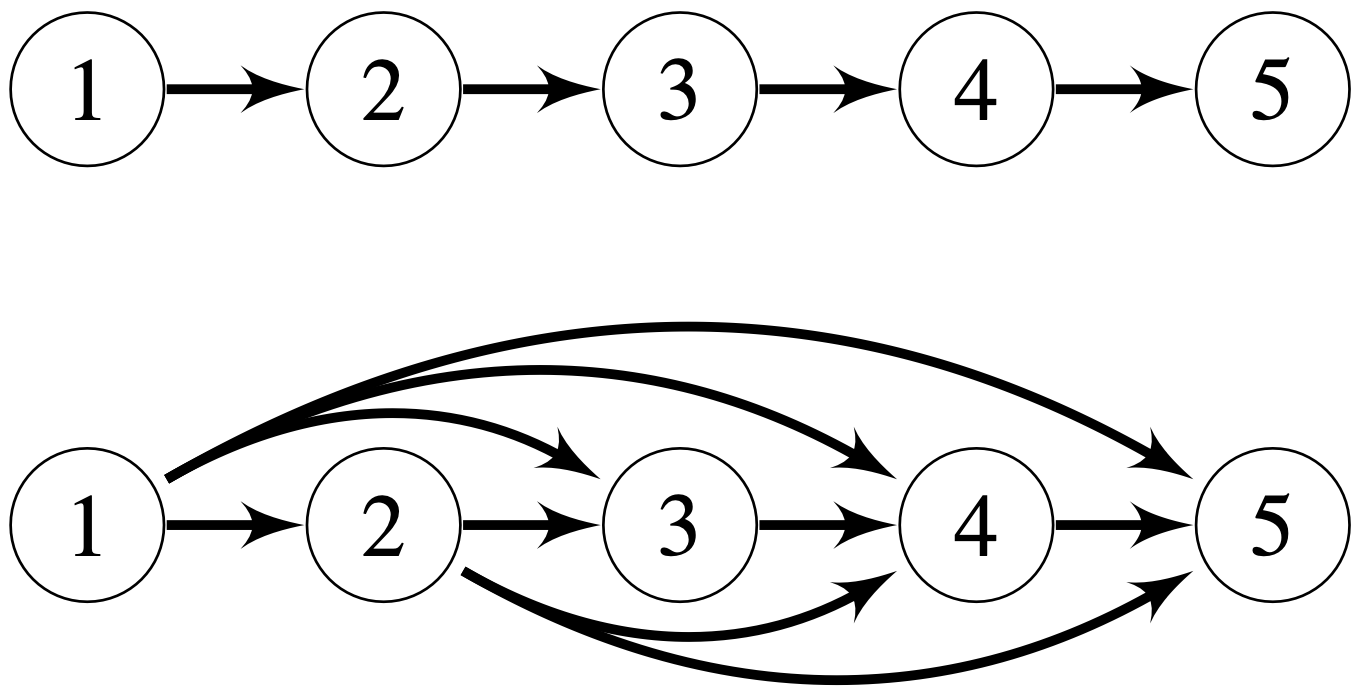} &
        \includegraphics[width=0.2\textwidth]{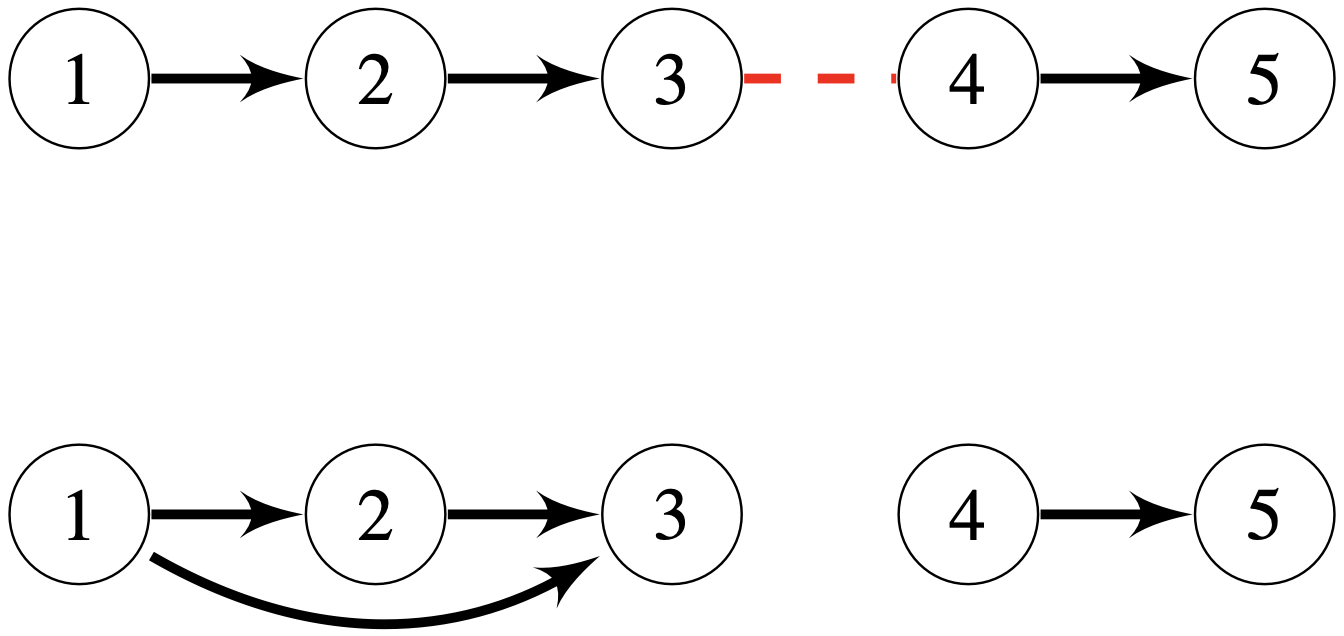} \\
        \small (a) Information graph $\mathcal{G}_I$ when the chain is fully connected. &
        \small (b) Information graph $\mathcal{G}_I$ when the chain is disconnected.
    \end{tabular}
    \medskip
    \caption{{\small
      Examples of information sharing topologies. The disconnected edge, indicating failure of message delivery, is shown by red arrows on the message-passing graph.}
    }
    \label{fig:cases}
\end{figure}

To set the stage to address Problem~\ref{prob:optimality_gap_prob}, it is pivotal to first examine the consequences of failed message deliveries and the ensuing disruption in the flow of information during the algorithm's operation. For this purpose, we use two specific graph topologies: the communication graph $\mathcal{G}$, which describes the sequence of message transfers, and the information graph $\mathcal{G}_I$, which describes the dissemination of information as a result of the message-passing process; see Fig.~\ref{fig:cases}. In the information graph, an arrow from agent $i$ to agent $j$ indicates that agent $j$ has successfully received information from agent $i$, either directly or through message-passing by other agents preceding agent $i$. 

When implementing Algorithm 2, a fully successful message transmission sequence, depicted in Fig.~\ref{fig:cases}(a), equates to the formation of a complete graph in the undirected version of the communication graph $\mathcal{G}_I$. Conversely, interruptions in the message-passing path, as shown in Fig.~\ref{fig:cases}(b), result in an undirected version of $\mathcal{G}_I$ that lacks full connectivity. To evaluate the degree to which $\mathcal{G}_I$ approaches a complete graph configuration, we employ the \emph{clique number} of the information graph, denoted by $\mathcal{W}(\mathcal{G}_I)$, as a metric. This measure, indicative of the size of the largest clique (complete sub-graph) within the graph, serves to quantify the graph's connectivity level. In the context of Algorithm~\ref{alg:sequential_decentral}, a complete $\mathcal{G}_I$ yields a clique number equal to $n$. Interruptions in message passage, however, diminish this number, signifying reduced information connectivity. One can expect that the larger the clique number, indicating a larger fully connected component in the information graph, the better the optimality gap should be for the sequential greedy algorithm. This hypothesis has been formally validated in~\cite{Gharesifard1}, which demonstrates that the optimality gap of a sequential greedy algorithm tackling problem~\eqref{eq::mainProblem}--under the condition that agents' policy sets do not overlap--can be effectively quantified as follows. 

\begin{lem}\longthmtitle{Optimally gap under deterministic unsuccessful message-passing~\cite{Gharesifard1}}\label{lem::deter_gap}
\emph{Consider the optimization problem~\eqref{eq::mainProblem}. The optimality gap of the sequential greedy algorithm when some message-passing paths are disconnected is given by} $ f(\bar{\mathcal{S}}_{SG}) \geq \frac{1}{2+n-\mathcal{W}(\mathcal{G}_I)} f(\mathcal{S}^\star).$\boxend
\end{lem}
\smallskip
This bound recovers the well-known optimality gap of $1/2$ when the information graph is complete, i.e., $\mathcal{W}(\mathcal{G}_I)=n$.

\smallskip

\begin{figure}[t]
    \centering
    \includegraphics[width=0.48\textwidth]{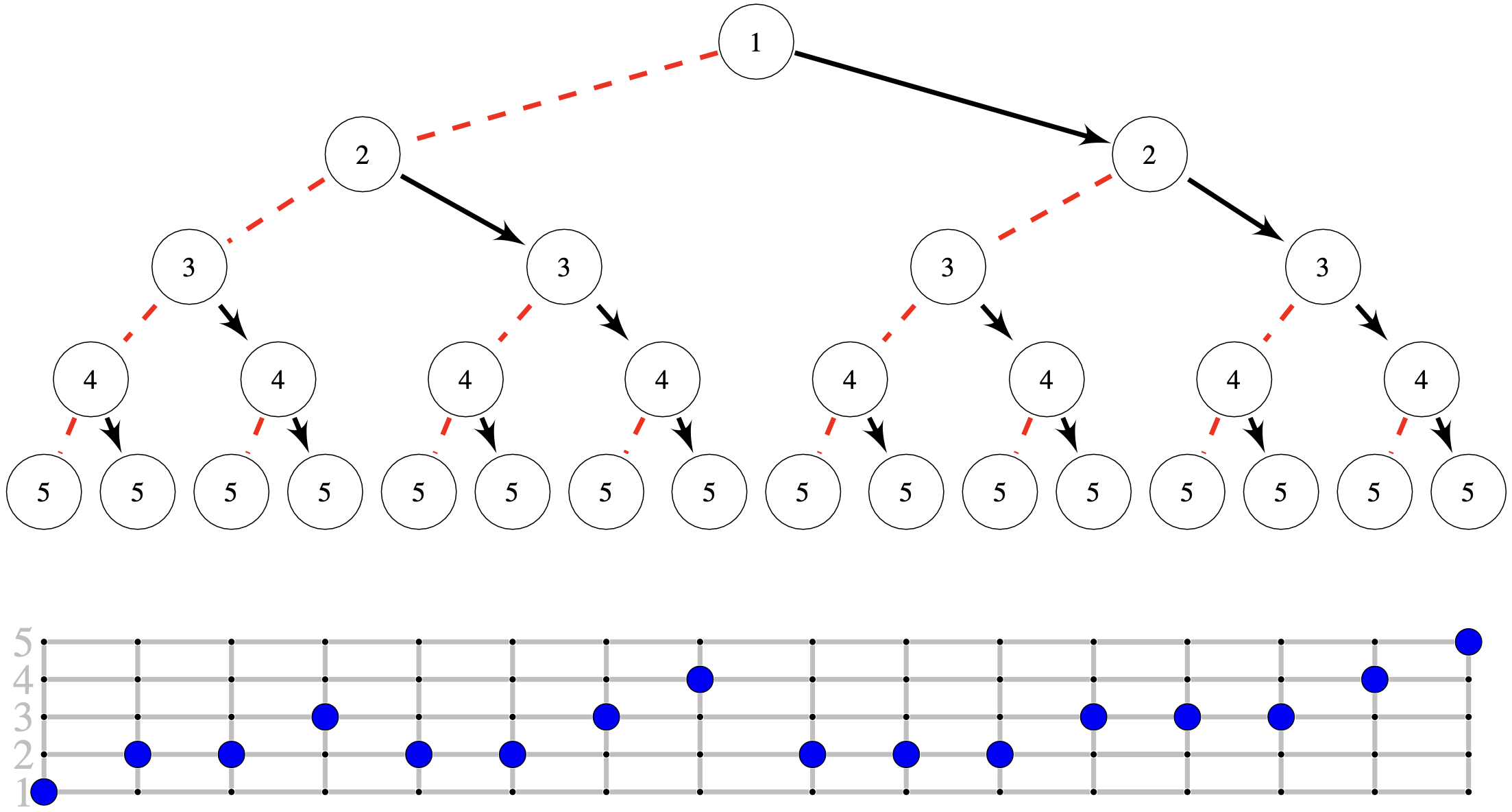}
\caption{{\small
  Examples of information sharing topologies. The top plots show the communication graph $\mathcal{G}$ along with its consequent maximum clique number. Arrow going from agent $i$ to agent $j$ means that agent $j$ receives agent $i$'s information. In red are represented the communication failures.}
 }
\label{fig:decision_tree}
\end{figure}

When message-passing success is probabilistic, both the outcome of the algorithm's execution and the clique number of the information graph become random variables. To deduce the expected optimality gap, we propose employing a decision tree representation to account for all possible message-passing outcomes. The upper diagram in Fig.~\ref{fig:decision_tree} shows this decision tree for the network of five agents shown in Fig.~\ref{fig:cases}, where the red dashed lines represent unsuccessful messages and the black arrows denote successful ones. The lower diagram in Fig.~\ref{fig:decision_tree} displays the associated clique number for each message-passing sequence outcome. For instance, the leftmost path on the decision tree, indicating a scenario where no messages are successfully passed, results in a clique number of one. In contrast, the rightmost path, which signifies a scenario in which all messages are successfully delivered, resulting in $\mathcal{W}(\mathcal{G}_I)\!=\!n\!=\!5$.
With a comprehensive overview of all possible message-passing scenarios,the expected optimality gap of Algorithm~\ref{alg:sequential_decentral} can be accurately characterized as follows.  

\begin{thm}\longthmtitle{Optimal Gap under Probabilistic message-passing}\label{thm::prob_gap} \emph{Regarding Problem~\ref{prob:optimality_gap_prob}, the probabilistic optimality gap $\alpha_p$ in~\eqref{eq:expected_optim_gap} can be expressed as
\begin{equation}\label{eq::alpha_p} \alpha_p = \sum\nolimits_{l=1}^{n} \frac{1}{2+n-l} P(\mathcal{W}(\mathcal{G}_I)=l), \end{equation} where $P(\mathcal{W}(\mathcal{G}_I)=l)$ denotes the probability that the clique number of the information graph $\mathcal{G}_I$ is $l$. }\end{thm}
\begin{proof} 
Since $\alpha_p$ is the expected probabilistic optimality gap, we calculate it by summing over all the possible optimality gaps computed from Lemma~\ref{lem::deter_gap} based on the clique numbers $l$ within the range from $1$ to $n$, each weighted by its probability $P(\mathcal{W}(\mathcal{G}_I)=l)$. \end{proof}

As shown in Fig.~\ref{fig:decision_tree}, a probabilistic communication chain of length $n-1$ has $2^{n-1}$ possible message-passing sequences. Denote the collection of all such sequence outcomes as $\mathcal{S}_{seq}$, and the random variable associated with any given sequence outcome as $seq$. The probability of each $seq\in\mathcal{S}_{seq}$ can be computed from a combination of $n-1$ Bernoulli distributions, each representing a possible connection between agents, that is, $seq \sim \mathsf{B}(p_1, \ldots, p_{n-1})$. Consequently, computing the clique number and associated probability for each sequence individually, will lead to an exponential time computation to determine $\alpha_p$. However, as seen in Fig.~\ref{fig:decision_tree}, multiple sequence outcomes can have the same clique number. In what follows, we propose a methodology that groups sequences based on their underlying structure. Specifically, to determine the probability of each clique number, we analyze the structure common to sequences that result in the same clique numbers and derive an exact formula for these probabilities. Our approach reduces the complexity of computing $\alpha_p$ to polynomial time. The critical information in our study is that the clique number is defined by the length of the longest directed path in the message-passing graph, plus one. We start with the following result. In this result and what follows, to simplify the notation, we write $P(\mathcal{W}(\mathcal{G}_I))$ as $P(\mathcal{W})$.
\begin{lem}\longthmtitle{Probability of Achieving a Specific Clique Number in Probabilistic Sequential Message-Passing}\label{lem::max_clique_prob} 
\emph{The probability that message-passing sequences successfully achieve a clique number of $l$ is given by
{\small
\begin{equation*}
    P(\mathcal{W}\!=\!l) \!=\! P(\mathcal{W} \!\geq\! l)-P(\mathcal{W} \!\geq\! l+1).
\end{equation*}}
}\end{lem}
\medskip
\begin{proof} 
{\small$P(\mathcal{W} \!\geq\! l)$} represents the probability that message-passing sequences achieve a clique number that is at least $l$. These sequences, which satisfy $\mathcal{W}(\mathcal{G}_I) \geq l$, include those with exact clique numbers ranging from $l$ to $n$, and these sequences are mutually exclusive. Thus, we express {\small\begin{align*} P(\mathcal{W} \geq l) = P(\mathcal{W}=l)+P(\mathcal{W}=l+1)+ &\cdots+P(\mathcal{W}=n). \end{align*} } Similarly, {\small$P(\mathcal{W} \!\geq\! l+1) \!=\! P(\mathcal{W}=l+1)+\cdots+P(\mathcal{W}=n)$}. The proof then follows by deducting $P(\mathcal{W} \geq l+1)$ from $P(\mathcal{W} \geq l)$. 
\end{proof}

\begin{figure}[t]
    \centering
    \begin{tabular}{c} 
        \includegraphics[width=0.45\textwidth]{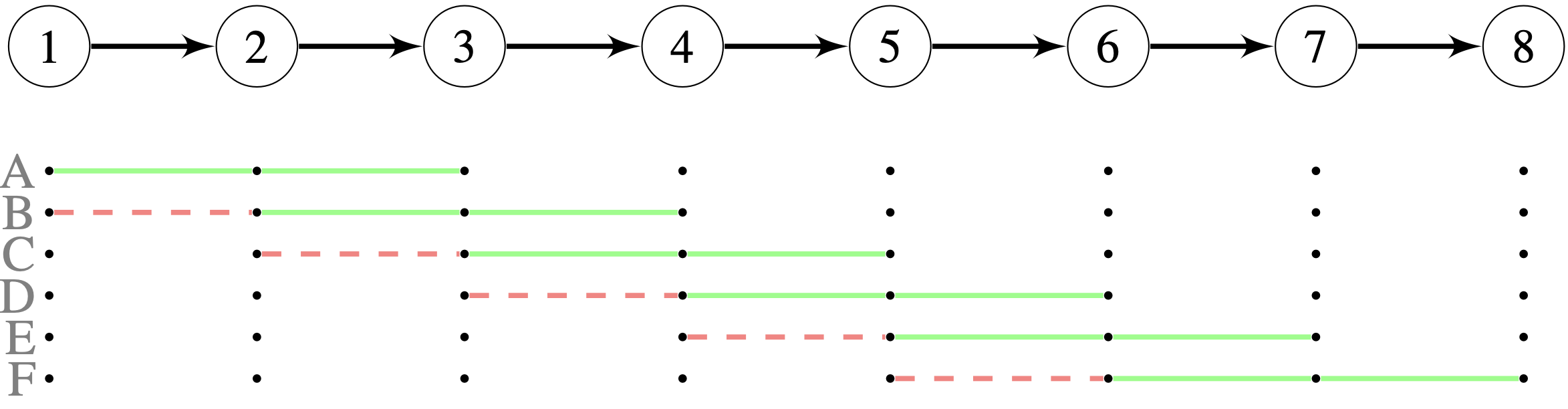} \\
        \small (a) Generative sequences of size $l=2$. \\[10pt] 
        \includegraphics[width=0.45\textwidth]{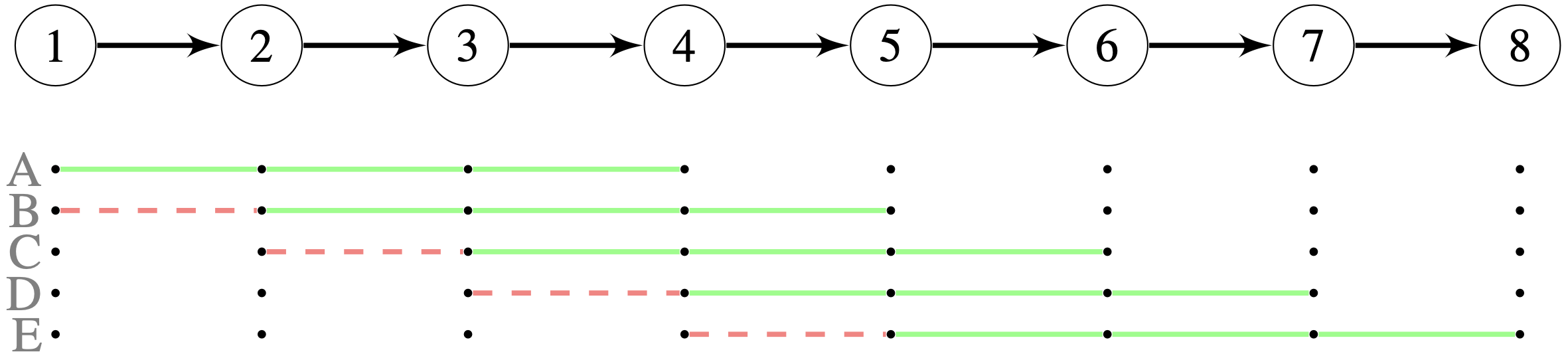} \\
        \small (b) Generative sequences of size $l=3$. 
    \end{tabular}
    \medskip
    \caption{{\small
    Examples of generative sequences, which generate families of message-passing sequences that have connected components of at least length two (top plot) and length three (bottom plot). A green link indicates an imposed connected edge, and a red link indicates an imposed disconnected edge. For example, generative sequence C in the top plot corresponds to $g(\{3,4\},\{2\})$ and in the bottom plot corresponds to $g(\{3,4,5\},\{2\})$.}}
    \label{fig:sequences}
\end{figure}

In what follows, we explain how to compute $P(\mathcal{W} \geq l+1)$ to fully describe $\alpha_p$ from~\eqref{eq::alpha_p}. 
$P(\mathcal{W} \geq l+1)$ is equivalent to determining the probability of a family of message-passing sequences achieving a maximum connected component length of $l$. To this end, we introduce some essential definitions. A \emph{family of message-passing sequences}, denoted by $\mathcal{F}$, refers to a group of sequences that exhibit a specific pattern of connectivity.
Thus, a \emph{family of message-passing sequences} is an \emph{event} associated with the random variable $seq$.
With a slight abuse of terminology, we define a \emph{generative sequence} as the guideline in which a connectivity structure is imposed on some of the edges of the massage-passing sequence, e.g., the first two edges are connected, but the fourth is not. Thus, a generative sequence becomes \emph{basis} for a family of sequences in which all members have the first two edges connected but the fourth not. We denote a generative sequence by $g(\mathcal{C},\mathcal{D})$, where $\mathcal{C},\mathcal{D}\subset\{1,\cdots,n-1\}$, respectively, specify the set of connected edges and disconnected edges of the generative sequence. For examples of generative sequences, see Fig.~\ref{fig:sequences}. We let $\mathcal{F}(g(\mathcal{C},\mathcal{D}))$ to denote the family generated by $g(\mathcal{C},\mathcal{D})$. 

\begin{lem}[Probability of a family]\label{eq::fam_g_prob}
\emph{  Given a massage passing sequence of $n$ agents with the corresponding probability of successful message-passing of $\{p_1,p_2,\cdots,p_{n-1}\}$, the probability of family generated by $g(\mathcal{C},\mathcal{D})$ is}
\vspace{-5pt}
\begin{align}\label{eq::p_generative}
    P(\mathcal{F}(g(\mathcal{C},\mathcal{D}))=\prod\nolimits_{i\in\mathcal{C}}p_i\prod\nolimits_{j\in\mathcal{D}}(1-p_j).
\end{align}
\end{lem}
\smallskip
Lemma~\ref{eq::fam_g_prob}'s proof follows trivially by writing all the outcomes generated by $g(\mathcal{C}, \mathcal{D})$ and adding their probabilities.

We call two generative sequences \emph{independent} when the families they initiate are distinct, in the sense that there exists no $seq \in \mathcal{S}_{seq}$ that is a member of the families generated by these generative sequences. For example, in top plot of Fig.~\ref{fig:sequences} generative sequences A and B, and A and C are independent, but generative sequences A and D are not independent.  

Next, we walk through the process of creating a family of message-passing sequences that feature connected components with a minimum length of $l \in \{1, \ldots, n-1\}$. For a specified $l$, we consider the generative sequences $g_i^l(\mathcal{C}_i, \mathcal{D}_i)$, where $i \in \{1, \ldots, n-l\}$, with $\mathcal{C}_i = \{i, i+1, \ldots, i+l-1\}$ representing the pre-specified connected components, and $\mathcal{D}_i = \{i-1\}$ indicating the pre-specified disconnected components, noting that $\mathcal{D}_1 = \emptyset$. These generative sequences each contribute to the creation of families of message-passing sequences that ensure the presence of connected components at least $l$ units long. Collectively, these sequences constitute $\cup_{i=1}^{n-l}\mathcal{F}(g^l_i)$, capturing the full set of message-passing sequences with connected components of at least $l$ in length, as depicted in Fig.~\ref{fig:sequences} through examples involving a chain of $8$ agents. It is noteworthy that
\begin{align} \label{eqn::dep_probability} P(\mathcal{W}\!\geq\!l+1) = P(\cup_{i=1}^{n-l}\mathcal{F}(g_i^l)). \end{align}
Given the construction where $\mathcal{C}_i \cap \mathcal{D}_{j+1} \neq {\emptyset}$ for any $i \in \{1, \ldots, l\}$ and $j \in \{i, \ldots, l+1\}$, as illustrated in Fig.~\ref{fig:sequences}, the families generated by $g_i^l$ for $i \in \{1, \ldots, l+1\}$ are independent and thus mutually exclusive events. Therefore,
$$P(\cup_{i=1}^{l+1}\mathcal{F}(g^l_i)) = \sum\nolimits_{i=1}^{l+1}P(\mathcal{F}(g^l_i)).$$

However, families from $i=l+1$ to $i=n-l$ have dependencies that we discuss next. With the same statement used for $i\in\{1,\cdots,l+1\}$, a family $\mathcal{F}(g_i^l)$ with $i\in\{l+2,\cdots,n-l\}$ will be dependent with families with $j\in\{1,\ldots,i-l-1\}$ as $(\mathcal{C}_i\cup\mathcal{D}_i)\cap(\mathcal{C}_j\cup\mathcal{D}_j)\!=\!\emptyset$. Therefore, we are seeking for the independent set of families in $\mathcal{F}(g_i^l)$ which can defined as the sequences in $\mathcal{F}(g_i^l)$ that do not belong to previous families $\mathcal{F}(g_j^l)$, in set notation $\mathcal{F}(g_i^l)\backslash\mathcal{F}(g_j^l)$. Note that by invoking a general probability rule \footnote{For two independent events $A$ and $B$, we have $P(A\backslash B)=P(A)\cdot(1-P(B))$\cite{8966}.} we obtain
\begin{equation} \label{eq::probab_dependent}
 P(\mathcal{F}(g_i^l)\backslash\mathcal{F}(g_j^l)) =  P(\mathcal{F}(g_i^l))\cdot(1- P(\mathcal{F}(g_j^l)))
\end{equation}
As a generalization for the previous expression, for the sake of understanding, let us consider the families that creates dependency to $\mathcal{F}(g_i^l)$, $\cup_{j=1}^{i-l-1}\!\mathcal{F}(g_j^l)$, such as $\mathcal{S}_i$. Then, notice~that
\begin{equation*}
\begin{split}
P(\mathcal{F}(g_i^l)\backslash\mathcal{S}_i) &= P(\mathcal{F}(g_i^l))\cdot(1-P(\mathcal{S}_i)) \\
&= P(\mathcal{F}(g_i^l))\cdot(1-P(\cup_{j=1}^{i-l-1}\mathcal{F}(g_j^l))) \\
&=P(\mathcal{F}(g_i^l))\cdot(1- \sum\nolimits_{j=1}^{i-l-1}P(\mathcal{F}(g_j^l))).
\end{split}
\end{equation*}

To obtain a close expression for \eqref{eqn::dep_probability} it is needed to find the independent expression for each family. By a simple set operation it can be seen that 
\begin{equation*}
 \begin{split}
  \cup_{i=1}^{n-l}\mathcal{F}(g_i^l) &= (\cup_{i=1}^{l-1}\mathcal{F}(g_i^l))\cup(\cup_{i=l}^{n-l}\widetilde{\mathcal{F}}(g_i^l)) \\
  &=(\cup_{i=1}^{l-1}\mathcal{F}(g_i^l))\cup(\cup_{i=l}^{n-l}\mathcal{F}(g_i^l)\backslash\mathcal{S}_i)
 \end{split}
\end{equation*}
where $\widetilde{\mathcal{F}}(g_i^l)\!=\!\mathcal{F}(g_i^l)\backslash\mathcal{S}_i$ is the independent expressions for families $i\in\{l+2,\ldots,n-l\}$. Therefore
{\small
\begin{equation}\label{eqn:final_cdf}
  \begin{split}
      &P(\mathcal{W}\!\geq\! l+1)=P(\cup_{i=1}^{n-l}\mathcal{F}(g_i^l)) \\
      &=P((\mathcal{F}(g^l_1),\cdots,\mathcal{F}(g^l_{l-1}))\cup(\mathcal{F}(g^l_{l})\backslash\mathcal{S}_l,\cdots,\mathcal{F}(g^l_{n-l})\backslash\mathcal{S}_{n-l}))) \\
      &=\! \sum\nolimits_{i=1}^{l+1} P(\mathcal{F}(g_i^l)) + \sum\nolimits_{i=n\!-\!l}^{l} P(\mathcal{F}(g_i^l)\backslash\mathcal{S}_i) \\
      &=\! \sum\nolimits_{i=1}^{l+1}\!\! P(\mathcal{F}(g_i^l))+\! \sum\nolimits_{i=l+2}^{n-l}\!\! P(\mathcal{F}(g_i^l))\!\cdot\!(1\!-\!\sum\nolimits_{j=1}^{i-l-1}\!\!P(\mathcal{F}(g_j^l))).\!
  \end{split}
\end{equation}
}
Each $P(\mathcal{F}(g_j^l))$ is computed from~\eqref{eq::p_generative}. Notice that equation \eqref{eqn:final_cdf} can be computed in $\mathcal{O}(n-l)$ oracle calls to~\eqref{eq::p_generative} and there is no need of computing all of the $2^{n-1}$ the possible sequences. With equation~\eqref{eqn:final_cdf} at hand, from Lemma~\ref{lem::max_clique_prob} we arrive at 
\begin{equation}
 \label{eqn::real_clique_prob}
 P(\mathcal{W}=l) = P(\cup_{i=1}^{n-l}\mathcal{F}(g_i^l)) - P(\cup_{i=1}^{n-l-1}\mathcal{F}(g_i^{l+1})).
\end{equation}
Now, using expression~\eqref{eqn::real_clique_prob},  the proposed optimality gap in Theorem~\ref{thm::prob_gap} can be easily computed in polynomial time. That not only allows the computation of the gap itself, but creates the possibility of pursuing an analysis of how the algorithm would perform in the worst-case scenario as a function of the parameters of the system.

\section{Extra resources assignment effect on the probabilistic gap}
\label{sec::bernoulli}
\begin{figure*}[t]
    \centering
    \begin{tabular}{@{\hspace{15pt}}  p{0.3\textwidth} @{\hspace{15pt}} p{0.3\textwidth} @{\hspace{15pt}} p{0.3\textwidth}}
        \includegraphics[width=4cm,height=3.5cm]{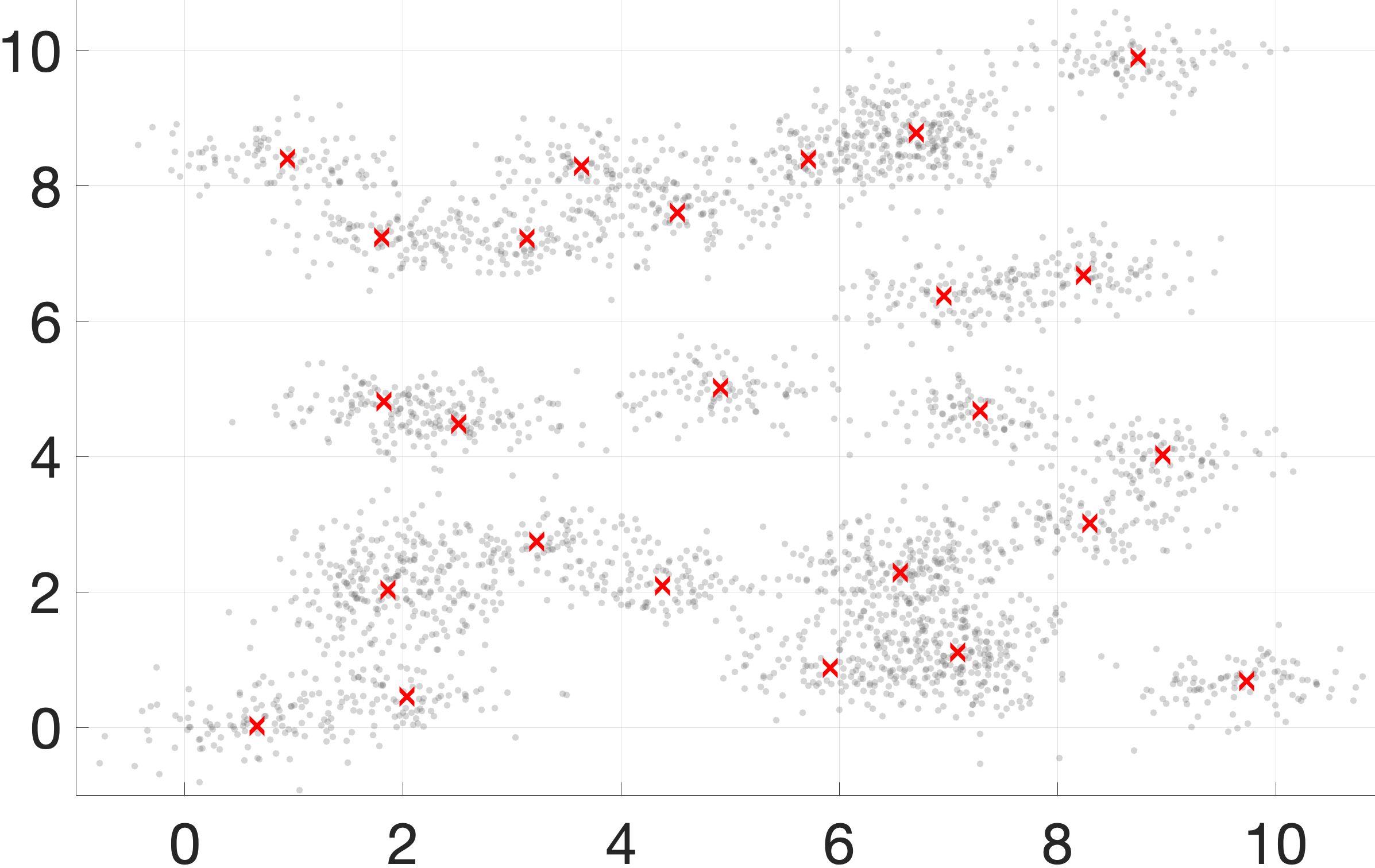} &
        \includegraphics[width=4cm,height=3.5cm]{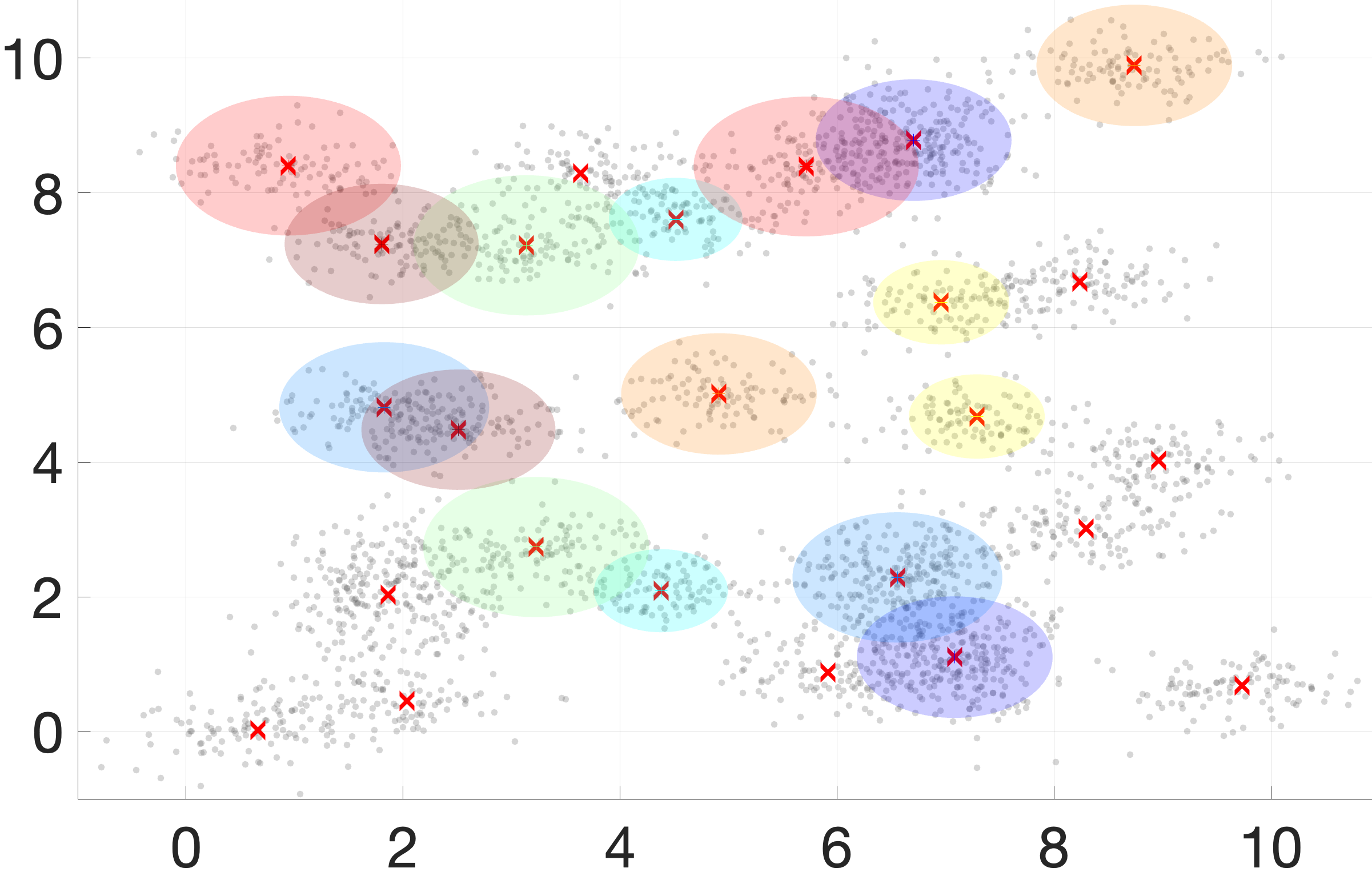} &
        \includegraphics[width=4cm,height=3.5cm]{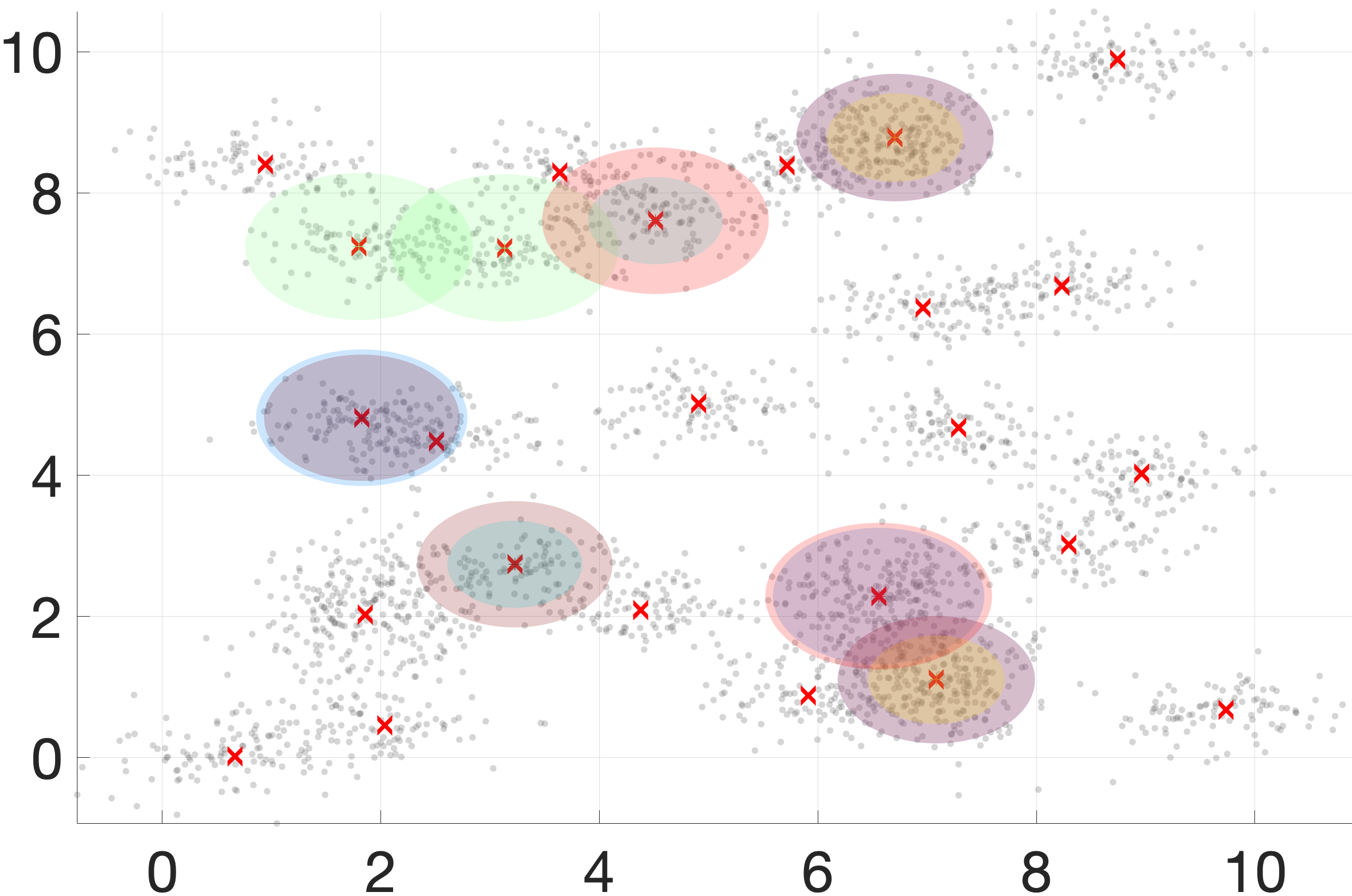}
    \end{tabular}
    \vspace{-0.1in}
    \medskip
    \caption{{\small Coverage Problem: (from left to right) initial distribution, coverage considering under no communication failure and coverage under the outcome $[0,1,1,1,0,1,1]$ of probabilistic communication. The disks show the coverage footprint of the sensors; the sensors corresponding to each agent is colored similarly. Note that when communication chain breaks, some agents tend to go to the same placement, reducing the coverage substantially; in the fully successful communication we achieve a solution $f(\mathcal{S})=1879$, while in the disconnection situation $f(\mathcal{S})=1246$.}}
    \vspace{-10pt}
    \label{fig::example}
\end{figure*}

By allowing repeated communication, we can increase the probability of message delivery and subsequently improve the optimality gap. However, multiple message-passing increases the end-to-end delay of executing the algorithm and may not be possible in resource limited scenarios. 
Therefore, the natural research question arises pertaining to the allocation of a finite quantity of multiple message-passing rights in an optimal manner. 
\begin{prob}[multiple message-passing right allocation]  \label{prob::multiple_message}
Given the decentralized sequential greedy Algorithm~\ref{alg:sequential_decentral}, let $p_i$ be the probability that the message $\bar{\mathcal{S}}_i$ from agent $i \in \mathcal{A}$ reaches its out-neighbor $i+1$. If only one agent can transmit its message twice, which agent should be chosen to maximize $\alpha_p$ in~\eqref{eq:expected_optim_gap}?\boxend
\end{prob}
\smallskip
The key observation is that the best agent to reinforce is not solely the one with the lowest communication probability but a combination of communication reliability and the agent's position in the communication chain. This is demonstrated in our empirical study in Section~\ref{sec::numerical}.

The message-passing process from agent $i$ to agent $i+1$ follows a Bernoulli distribution $$
    p_i(T_i,r) = \frac{T_i!}{r!\,(T_i-r)!} \cdot p_{i,0}^r \cdot (1-p_{i,0})^{T_i-r},$$ where $p_{i,0}$ is the probability of agent's successful delivery in one communication, $T_i$ is the number of trials (messages sent) and $r$ is the number of successes (messages delivered). Since communication between two nodes is established when at least one success occurs, the probability that the recipient gets the message is \begin{equation} p_i(T_i,r \!\geq\! 1) = 1 - p_i(T_i,r = 0)= 1 - (1 - p_{i,0})^{T_i}.\end{equation}
Note that $p_i(T_i,r\geq 1)$ is the new communication probability from agent $i$ to agent $i+1$, for the sake of understanding, we will keep referring to them as $p_i$. By considering a Bernoulli measure, expression in Lemma~\ref{eq::fam_g_prob} can be re-defined as 
$$P(\mathcal{F}(g(\mathcal{C},\mathcal{D}))=\prod\nolimits_{i\in\mathcal{C}}(1-(1-p_i)^{T_i})\prod\nolimits_{j\in\mathcal{D}}(1-p_j)^{T_j},$$
enabling us to analyze the behavior of message-passing reinforcement for $T_i\geq1$. Note that when the message-passing count $T_i$ for a device $i$ increases, the expected optimality gap needs to be recalculated. As previously mentioned, for each $l$, calculating its probability $P(\mathcal{W}\geq l)$ requires a time complexity of $\mathcal{O}(n-l)$. Consequently, the gap $\alpha_p$ can be determined within $\mathcal{O}(\frac{n\cdot(n-1)}{2})$, indicating that optimizing the communication strategy is a polynomial-time process. Further examination of \eqref{eqn::real_clique_prob} reveals that doubling the message-passing trail for agent $k$ does not influence families where $i\leq k-l+1$. Thus, by equipping the system with a certain memory capacity, computation time of $\alpha_p$ for all possible reinforcements can be significantly reduced.

\section{Empirical study of the probabilistic gap}
\label{sec::numerical}
Consider a multi-sensor deployment problem where a group of $8$ heterogeneous agents $\mathcal{A}=\{1, \ldots, 8\}$, each with two sensors to deploy at some prespecified deployment points of interest. There are $25$ total locations where sensors can be deployed, but each agent $i\in\mathcal{A}$ has access to only a subset of $12$ locations, denoted by $\mathcal{B}_i$. The sets $\mathcal{B}_i$ are not distinct, but using a simple trick, we can create the local selection set of the agents in a distinct way as $\mathcal{P}_i = \{(i, b) \mid b \in \mathcal{B}_i\}$ for $i \in \mathcal{A}$. The environment includes a set of $2,200$ randomly generated sampled points, denoted by $\mathcal{V}$. The sensor deployment objective is to cover as many points from $\mathcal{V}$ as possible, which is accomplished by solving a maximization problem of the form~\eqref{eq::mainProblem} with $\kappa_i=2$ and the utility function given by $ f(\mathcal{S}) = \sum\nolimits_{p \in \mathcal{V}} g(p)$, where $g(p)=1$ if there exists at least one element $(\ell, b) \in \mathcal{S}$ such that $\|c - p\| \leq \rho_{\ell}$; otherwise $g(p)=0$. Here, $\rho_{\ell}$ is the coverage radius of sensor $\ell$. This utility function is known to be submodular and monotone increasing. This problem is illustrated in Fig.~\ref{fig::example}.

\begin{figure}[t]
\vspace{2mm}
\centering
\includegraphics[width=0.45\textwidth]{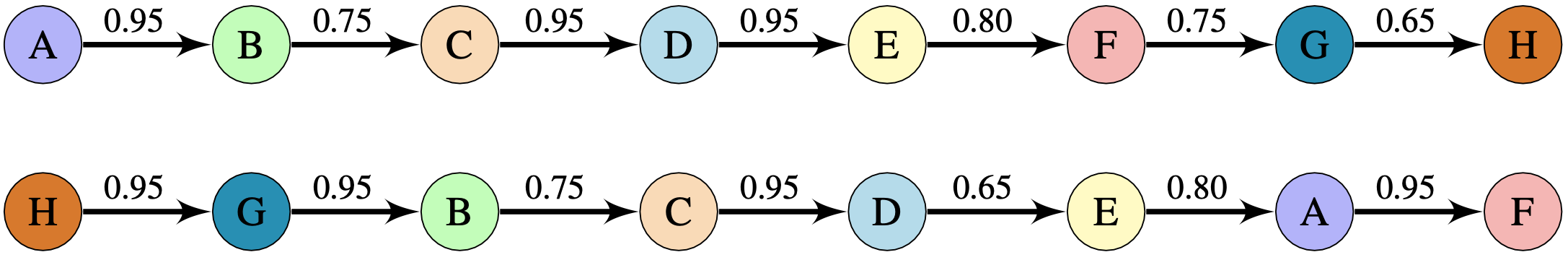}
\caption{{\small
  Two different possible $m$ communication chains.}}
  \vspace{8pt}
\label{fig:example_sequences}
\end{figure}
 
For visualization of the communication chain, we label the agents alphabetically. Now consider two different sequences of communication: lexicographic order \textit{ABCDEFGH} and a random shuffle \textit{DBHGFCAE}, as shown in Fig.~\ref{fig:example_sequences}. Note that with $n$ agents, there are up to $n!$ possible communication chains that can be established. In this section, we consider only two sequences to illustrate the particular behavior under the allowance for extra message-passing. It is important to note that due to the probabilistic nature of the problem, the results presented in this section are the mean values from 10,000 iterations for each scenario.

Table~\ref{tab:results} shows the results of our study for the communication chains illustrated in Fig.~\ref{fig:example_sequences}. The second column presents the average utility value over 10,000 iterations when agents use Algorithm~\ref{alg:sequential_decentral} to determine their deployment locations. The third column shows $\alpha_p$ computed from~\eqref{eq::alpha_p}. The fourth column identifies the agent $a^\star$ whose communication is reinforced by allowing two back-to-back communication trials. The agent to be reinforced is determined by computing $\alpha_p$ for all possible reinforcements, with these values shown in Table~\ref{tab:enhancement}. The fifth column indicates the edge label that is reinforced (the link coming out of $a^\star$). Finally, the sixth and seventh columns present the average utility value and $\alpha'_p$ computed from~\eqref{eq::alpha_p} under communication reinforcement. As expected, we observe that $\alpha'_p > \alpha_p$, indicating that reinforcement improves the expected optimality gap. This improvement is also reflected in the utility values. Another noteworthy observation is that the reinforced agent is not necessarily the one with the lowest communication probability, and the reinforced edge differs between sequences.

\begin{table}[t] 
    \centering 
     \caption{\footnotesize Optimality gap and utility value with and without communication reinforcement. The agent $a^\star$, which is allowed to communicate twice, is selected based on achieving the highest $\alpha_p$.}\vspace{-0.06in}
    \begin{tabular}{|c|c|c||c|c|c|c|}
        \hline
        \textbf{Sequence} & \textbf{${f}(\cdot)$} & $\alpha_p$ & \textbf{$a^\star$} & $e^\star$ & \textbf{${f}'(\cdot)$} & $\alpha_p'$ \\ \hline
        ABCDEFGH & 1324.21 & 0.2692 & F & 6 & 1610.88 & 0.3243  \\ \hline
        DBHGFCAE & 1554.54 & 0.2788 & C & 4 & 1749.21 & 0.3313 \\ \hline
    \end{tabular}
    \label{tab:results}
\end{table}
\begin{table}[t]
    \footnotesize
    \centering 
        \caption{{\footnotesize $\alpha_p$ vs. different agent $a\in\{1,\cdots,7\}$ allowed to communicate twice.}}\vspace{-0.07in}
    \begin{tabular}{|d{0.17}|d{0.06}|d{0.06}|d{0.06}|d{0.06}|d{0.06}|d{0.06}|d{0.06}|d{0.06}|}
        \hline
      \!\!\! \textbf{Sequence}  & 1 & 2 & 3 & 4 & 5 & 6 & 7\\ \hline
  \!\!\! ABCDEFGH\!\!\! &\!\!\!0.2974& \!\!\!0.3192 & \!\!\!0.3006 &\!\!\!0.3012 & \!\!\!0.3218 & \!\!\textbf{0.3243}\!\! & \!\!\!0.3030  \\ \hline
    \! \!\! DBHGFCAE\!\!  &\!\!\!0.2998 &\!\!0.3286\!\!& \!\!\!0.3028 &\!\!\!\textbf{0.3313} &\!\!\!0.3166& \!\!\!0.3218 & \!\!\!0.3004 \\ \hline
    \end{tabular}
    \label{tab:enhancement}
\end{table}

\section{Conclusions}
\label{sec:conclusions}
This paper addressed the problem of decentralized submodular maximization subject to partition matroid constraint using a sequential greedy algorithm with probabilistic inter-agent message-passing. We proposed a communication-aware framework that considers the reliability of communication between connected devices, emphasizing its impact on the optimality gap. Our analysis introduced the notion of the probabilistic optimality gap $\alpha_p$, highlighting its crucial role in understanding the expected performance of the sequential greedy algorithm under probabilistic communication. By characterizing $\alpha_p$ as an explicit function of communication probabilities, we created a framework to answer critical questions such as which agent should be reinforced (allowed multiple communications) to improve the optimality gap or how to compare the optimality gaps of various communication chains, where the order of agents in the chain changes while their reliability remains the same. In our empirical study, we specifically focused on the case where only one agent is allowed to communicate twice. However, the methodology we presented is generalizable and can be extended to scenarios where multiple agents are allowed multiple message-passing opportunities. This extension can be framed as a set function maximization problem subject to a uniform matroid constraint, with $\alpha_p$ as the utility function. What agents to reinforce can be efficiently decided using a sequential greedy approach, given that $\alpha_p$ is monotone increasing and, if shown to be submodular, would benefit from the known optimality gap of $0.63$. Future work will focus on formally proving the submodularity of $\alpha_p$ and developing strategies to optimize communication reinforcement in more complex scenarios involving multiple agents. We aim also to leverage insights from this study to design more robust and efficient decentralized systems for real-world applications, particularly in environments where communication reliability is variable and resources are limited.

\bibliographystyle{ieeetr}
\bibliography{main}

\end{document}